\documentclass[11pt,twoside]{article}
\usepackage{graphicx,epsfig,natbib,epstopdf}
\usepackage{CS18}
\begin{document}
%
%
%
\title{{\tt BinHab:}~A Numerical Tool for the Calculation of S/P-Type
Habitable Zones in Binary Systems}
%
%
\author{M. Cuntz, R. Bruntz}
\affil{Department of Physics, University of Texas at Arlington, Arlington, Texas, USA 76019;
Email: {\tt cuntz@uta.edu}, {\tt rbruntz@uta.edu}}
\begin{abstract}
The aim of this contribution is to introduce the numerical tool {\tt BinHab},
a publicly accessible code, available at The University of Texas at Arlington,
that allows the calculation of S-type and P-type habitable zones of
general binary systems.
\end{abstract}

\section{Introduction}

After several decades of research, planets\index{planet}
in stellar binary systems\index{binary system}
now constitute a well-established observational result.  Previous
examples include P-type orbits, when the planet is found to orbit
both binary components, as well as S-type orbits with the planet
orbiting only one of the binary components with the second component
acting as a perturbator; see, e.g., \cite{egg07} and \cite{neu07}
for selected observational results and data.

Another topic of significant importance, especially concerning the
astrobiological community, are studies of circumstellar and
circumbinary habitability\index{habitability}.
Previous work focused on the traditional concept of \cite{kas93}
and subsequent studies, where habitability is defined based on the
principal possibility that liquid water is able to exist on the
surface of an Earth-type planet\index{Earth-type planet} possessing
a CO$_2$/H$_2$O/N$_2$ atmosphere; for more sophisticated concepts
see, e.g., \cite{lam10}, and references therein.  Other relevant
investigations concern studies of
orbital stability\index{orbital stability}, especially
for (hypothetical) Earth-type planets in stellar
habitable zones\index{habitable zone}.
This type of work has been pursued for single as well as
multi-planetary\index{multi-planetary system} and
multi-stellar systems\index{multi-stellar system};
see, e.g., \cite{jon01},
\cite{nob02}, \cite{san07}, for early contributions as well
as, e.g., \cite{kan13} and \cite{jai14} for more recent work.
Some of these efforts resulted in stability catalogs of the
habitable zones of the planetary systems known at the time.

\cite{cun14a,cun14b} [Paper~I and II] forwarded a concise
approach for the investigation of habitable regions in stellar
binary systems, which forms the basis for {\tt BinHab}
(see Sect.~2).  In Sect.~3, we will give applications to
S/ST-type habitability for binaries of low mass stars.
Our conclusions and outlook are presented in Sect.~4.

\section{Description}

\subsection{Methods}

The method as conveyed has previously been given in Paper~I
and II; thus in the following we will focus on the most decisive
concepts, which include: (1) The consideration of a joint constraint
comprising orbital stability and a habitable region for a putative
system planet through the stellar radiative energy fluxes
(``radiative habitable zone"; RHZ) needs to be met.  (2) The treatment
of conservative, general and extended zones of habitability for the
various systems, referred to as CHZ, GHZ, and EHZ, respectively,
following the approach given by \cite{kas93} and subsequent work.
(3) The providing of a combined formalism, based on solutions of a
fourth-order polynomial, for the assessment of both S-type and P-type
habitability.  In particular, mathematical criteria are presented for
which kind of system S-type and P-type habitability is realized.

Following Paper~I, five different cases of habitability are identified,
which are: S-type and P-type habitability provided by the full extent
of the RHZs; habitability, where the RHZs are truncated by the additional
constraint of planetary orbital stability (referred to as ST and PT-type,
respectively); and cases of no habitability at all.  Regarding the
treatment of planetary orbital stability, the formulae of \cite{hol99}
are utilized.  {\tt BinHab} is suitable for both circular and elliptical
stellar binary components, the topic of Paper~II.  Figure~1 conveys
the flow diagram.  Note that for S-type orbits, the orbital stability
criterion operates as an upper limit of orbital stability (see Sect. 3),
whereas for P-type orbits, it operates as a lower limit of planetary
orbital stability.

{\tt BinHab} allows to consider general binary systems, including systems
containing evolved stars.  Surely, for main-sequence stars, there is an
intimate coupling between the input parameter T$_i$ and R$_i$ ($i=1,2$),
the stellar effective temperature and radius, on the one hand, and M$_i$,
the stellar mass, on the other hand, which is however not required for
the usage of the code\index{code}.  The simulations of Paper~I are largely
based on data given in \cite{gra05}.

\begin{figure}
\plotone{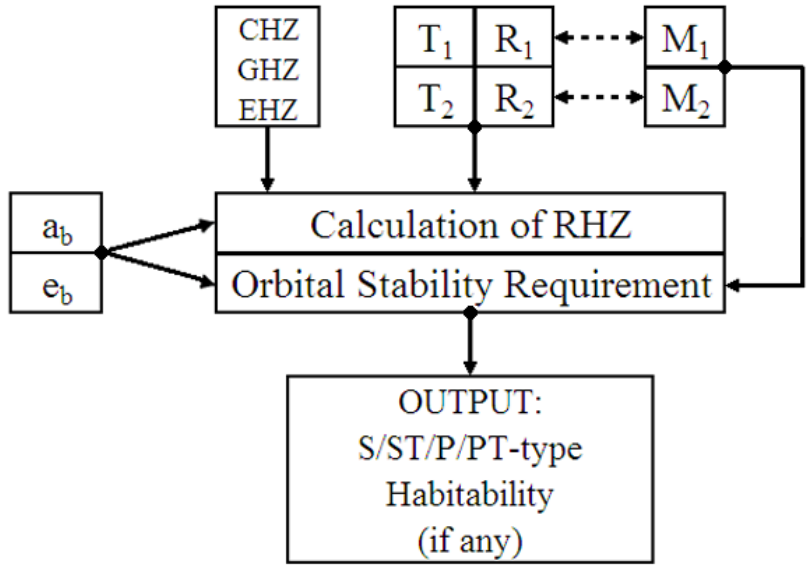}
\caption{Flow diagram of {\tt BinHab}}
\end{figure}

\subsection{Implementation}

To make {\tt BinHab} publicly accessible, we created a website through
which stellar system conditions could be entered, and the equations
pertinent to {\tt BinHab} would then be used to find whether any
habitable zone or zones exist, and if so, what type.  A virtual
server was set up and hosted by the UT Arlington IT department at
{\tt http://physbinhab.uta.edu}, as a dedicated server for the
{\tt BinHab} website. It allows the user to enter binary system
parameters (i.e., semi-major axis and eccentricity), separate
stellar parameters for the two stars (i.e., temperature, luminosity,
and mass), and the type of habitable zone the user would like to look
for (i.e., CHZ, GHZ, or EHZ).  Background information on the different
types of habitable zones is also given.

The output states whether there were any habitable zones found,
and if so, what type (S, ST, P, or PT) and the inner and outer
limit of any zones.  The website itself uses HTML
(HyperText Markup Language) to interface with the user, by
displaying information, accepting user input parameters, and
displaying the results of the {\tt BinHab} calculations. The
user's input is passed to PHP (PHP: Hypertext Preprocessor)
code, which also checks the validity of the input.  If any
of the parameters are out of range, the PHP code creates a
warning message, which is passed to the HTML code and
displayed to the user, explaining what needs to be adjusted.
If all parameters are acceptable, the PHP code saves them to
a new input file, runs the {\tt BinHab} Fortran binary on the
input file, and collects the output from the binary. The
output is then checked, formatted, and passed to the HTML
code, for display to the user.

The use of PHP allows for much more processing of the input
than is possible with HTML alone and is necessary for
running the Fortran binary. Fortran is used for implementing
the {\tt BinHab} algorithms, rather than PHP, for a variety
of reasons, including that it is much more widely used in
the scientific community and that as a compiled language,
it executes much faster than the interpreted PHP code.
Additionally, the Fortran code and input files are stored
in a segment of the server directory structure that is
inaccessible to the website, but accessible to the PHP code.
Hence, the Fortran code is not viewable or downloadable from
the web, as it can only be accessed by the PHP code, which
runs on the server, and therefore is not directly viewable
from the Internet.  In order to make the website easier to
find, we successfully submitted the URL to the two largest
search engines: Google and Bing (which also drives the Yahoo!
search engine).

\section{Example: Application to S/ST-Type Habitability for
Binaries of Low-Mass Main-Sequence Stars}

In order to illustrate the capacity of the method, we convey
studies of S/ST-type habitability for binaries of low-mass
main-sequence stars.  Here we focus on stars of masses
0.75~M$_\odot$, 0.65~M$_\odot$, and 0.50~M$_\odot$, corresponding
to spectral types of K2V, K6V, and M0V.  This type of work
is motivated by the observational finding that stars with masses
below about 0.8~M$_\odot$ constitute nearly 90\% of all stars in
the Milky Way\index{Milky Way} \citep[e.g.,][]{kro02,cha03}.
As an example we focus on
S-type and ST-type habitable zones in binaries consisting
of these types of stars.  In particular, we investigate the role
of the eccentricity of the binary system $e_{\rm b}$ on the width
of the S/ST-type habitable zones (if existing) for systems with
$2a_{\rm b} = 5.0$~AU as examples while focusing on results for
the GHZ (see Fig.~2).

The following aspects are identified: For all stellar mass
combinations, stellar habitable zones are found for eccentricities
below 0.20.  For the majority of models pursued, S-type habitability
is truncated due to the orbital stability requirement of the putative
planet resulting in ST habitability classification (see Table~1).
Stellar pairs with masses of 0.75~M$_\odot$ exhibit the broadest RHZs
due to their relatively high luminosities; however, in this case the
relatively large stellar masses lead to a significant truncation of
circumstellar habitability, resulting in relatively narrow habitable
zones.  On the other hand, stellar pairs with masses of 0.50~M$_\odot$ 
show S/ST-type habitable regions up to binary eccentricities of 0.65
(see Table~2), albeit the widths of the habitable zone are relatively
small compared to other primary--secondary mass combinations, especially
for cases of small binary eccentricities.  Starting at a well-defined
eccentricity, the width of the habitable zones decreases linearly as a
fraction of the binary eccentricity for all mass combinations, encompassing
both systems of equal and nonequal stellar masses, owing to the truncation
criterion \citep{hol99}.  Consistently smaller widths for the habitable
regions are identified for models based on CHZs relative to GHZs, as
expected.

\begin{figure}
\plotone{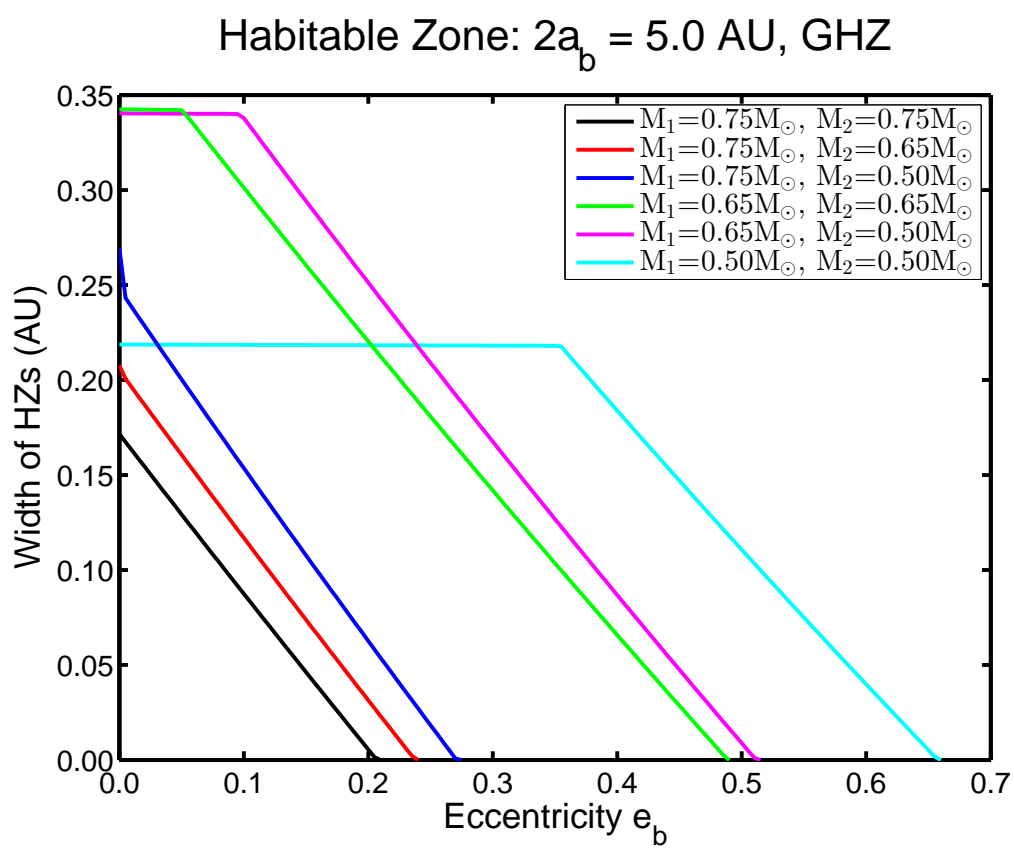}
\caption{Widths of S/ST-type habitable zones in low-mass binary systems.}
\end{figure}

\begin{table}
\caption{S/ST-type Habitability for Models of $2a_{\rm b} = 5.0$~{\rm AU}, GHZ}
\centering
\vspace{0.05in}
\vspace{0.05in}
\begin{tabular}{lccccccc}
\hline
\hline
\noalign{\vspace{0.03in}}
Model                                  &  0.0 &  0.1 &  0.2 &  0.3  &  0.4 &  0.5  &  0.6  \\
\noalign{\vspace{0.03in}}
\hline
\noalign{\vspace{0.03in}}
M$_1=0.75$~M$_\odot$, M$_2=0.75$~M$_\odot$ &  ST  &  ST  &  ST  &  ...  &  ... &   ...  &  ... \\
M$_1=0.75$~M$_\odot$, M$_2=0.65$~M$_\odot$ &  ST  &  ST  &  ST  &  ...  &  ... &   ...  &  ... \\
M$_1=0.75$~M$_\odot$, M$_2=0.50$~M$_\odot$ &  ST  &  ST  &  ST  &  ...  &  ... &   ...  &  ... \\ 
M$_1=0.65$~M$_\odot$, M$_2=0.60$~M$_\odot$ &  S   &  ST  &  ST  &  ST   &  ST  &   ...  &  ... \\
M$_1=0.65$~M$_\odot$, M$_2=0.50$~M$_\odot$ &  S   &  ST  &  ST  &  ST   &  ST  &   ST   &  ... \\
M$_1=0.50$~M$_\odot$, M$_2=0.50$~M$_\odot$ &  S   &  S   &  S   &  S    &  ST  &   ST   &  ST  \\
\noalign{\vspace{0.05in}}
\hline
\end{tabular}
\vspace{0.05in}
\end{table}

\begin{table}
\caption{Critical Values of $e_{\rm b}$ for S/ST-type Habitability}
\centering
\vspace{0.05in}
\vspace{0.05in}
\begin{tabular}{lcc}
\hline
\hline
\noalign{\vspace{0.03in}}
Model                                      &  CHZ   &  GHZ  \\
\noalign{\vspace{0.03in}}
\hline
\noalign{\vspace{0.03in}}
M$_1=0.75$~M$_\odot$, M$_2=0.75$~M$_\odot$ &  0.12  &    0.20  \\
M$_1=0.75$~M$_\odot$, M$_2=0.65$~M$_\odot$ &  0.15  &    0.23  \\
M$_1=0.75$~M$_\odot$, M$_2=0.50$~M$_\odot$ &  0.19  &    0.27  \\
M$_1=0.65$~M$_\odot$, M$_2=0.60$~M$_\odot$ &  0.43  &    0.48  \\
M$_1=0.65$~M$_\odot$, M$_2=0.50$~M$_\odot$ &  0.46  &    0.51  \\
M$_1=0.50$~M$_\odot$, M$_2=0.50$~M$_\odot$ &  0.62  &    0.65  \\
\noalign{\vspace{0.05in}}
\hline
\end{tabular}
\vspace{0.05in}
\end{table}

\section{Conclusions and Outlook}

We provided a short description of the features and capacities of the
numerical tool {\tt BinHab} hosted at The University of Texas at
Arlington. It considers a joint constraint including orbital stability
and a habitable region for a putative system planet through the
stellar radiative energy fluxes, among various other desirable
features.  Although the code has previously mostly been used to
investigate binary systems consisting of main-sequence stars
\cite{cun14a,cun14b}, it is highly flexible; it can also be
utilized for the calculation of habitable zones for systems
containing a subgiant, giant, or supergiant.  Concerning the
latter, {\tt BinHab} has already been used to study the habitability
of Earth-mass planets and moons in the Kepler-16 system \citep{qua12},
known to host a circumbinary Saturn-mass planet.  Ultimately, it is
our goal to expand the developed methods to multiple stellar systems,
which are of notable interest to the scientific community.

\acknowledgments{
This work has been supported in part by the SETI institute.  The authors
also acknowledge assistance by Zhaopeng Wang with computer graphics.
}

\end{document}